# Real-Time Fault Detection and Process Control Based on Multi-channel Sensor Data Fusion

Feng Ye, Zhijie Xia, Min Dai, Zhisheng Zhang
Department of Mechanical and Electrical Engineering
Southeast University
feye@seu.edu.cn

*Abstract*—Sensor signals acquired in the industrial process contain rich information which can be analyzed to facilitate effective monitoring of the process, early detection of system anomalies, quick diagnosis of fault root causes, and intelligent system design and control. In many mechatronic systems, multiple signals are acquired by different sensor channels (i.e. multi-channel data) which can be represented by high-order arrays (tensorial data). The multi-channel data has a high-dimensional and complex cross-correlation structure. It is crucial to develop a method that considers the interrelationships between different sensor channels. This paper proposes a new process monitoring approach based on uncorrelated multilinear discriminant analysis that can effectively model the multi-channel data to achieve a superior monitoring and fault diagnosis performance compared to other competing methods. The proposed method is applied directly to the high-dimensional tensorial data. Features are extracted and combined with multivariate control charts to monitor multi-channel data. The effectiveness of the proposed method in quick detection of process changes is demonstrated with both the simulation and a real-world case study.

## I. Introduction

With the development of smart sensor technologies, online monitoring is being widely applied in many industrial applications. The outputs of sensors are time-ordered data, which is also referred as waveform signal data or profile data [1,2]. Examples of such data include the force signals used to press seats into the engine head during an assembly process [3] and the power signals in the ultrasonic metal welding process [4]. The profile data acquired by sensors during the industrial process can provide rich information to develop an in-process quality control tool to monitor product quality and automatically control the process. The monitoring strategy is crucial during the process. For example, in the forging process, sensors are installed on the forging machine, which records the force exerting on dies.

There is a widespread study on modeling and monitoring of profile data [5-7]. For the purpose of profile monitoring and fault detection, Paynabar et al. [8] applied a nonlinear parametric regression model to develop a system which was robust and insensitive to the variations of temperature in production practice. A monitoring method which can adapt the parameters of the control chart automatically was designed by Grasso et al. [9]. Lei et al. [10] added up all channel profiles and applied principal component analysis (PCA) to the aggregated tonnage profiles to extract features. For the purpose of feature extraction, Bhattacharyya, Sengupta [11] proposed an adaptive sensor fusion method combined with signal processing techniques to extract features to estimate tool wear. Yang et al. [12] used both statistical features and wavelet features extracted from sensor signals to develop an adaptive learning method for tool wear estimation in the high-speed milling process.

Most of the researches focused on the analysis of individual profile (single signal). However, in many industrial processes, multiple signals are collected by different sensors. The outputs of these sensors are called multi-channel data which can be represented by high-order arrays (tensorial data). Multi-channel data can achieve better performance of signal monitoring. The rich data collected in the industry are leading to increasing demand for sensor fusion approaches to improve product quality. Multi-channel data is high-dimensional and have complex cross-correlation structure. Recently, some studies have been devoted to multi-sensor fusion in industrial processes. Grasso et al. [13] applied multi-way PCA to reduce the data dimensionality in order to improve the efficiency of the profile analysis system and monitored multi-channel profile data with control charts. Zerehsaz et al. [14] extracted a set of few monitoring features from the high dimensional profile data to detect tool wear. Paynabar et al. [15] applied uncorrelated multilinear principal component analysis (UMPCA) [16] for profile monitoring and fault diagnosis which accounted for the interrelationship of different profile channels. Compared to UMPCA which is based on PCA, linear discriminant analysis (LDA) is also a classical method for feature extraction. However, regular LDA cannot be operated on multi-channel data directly because it is a method for vectors. An uncorrelated multilinear discriminant analysis (UMLDA) [17] was proposed for face recognition and image processing. UMLDA operates on multidimensional data directly and extracts uncorrelated discriminant features by tensor-to-vector projection. Compared to the UMPCA algorithm which is unsupervised, UMLDA is a supervised multilinear feature extractor which will consider class information when extracting features and could be more suitable for face recognition. Although there is some exploratory research on the applications of UMLDA to face recognition and image processing, little research has been reported in the literature on using the UMLDA technique to analyze multi-channel data to automatically control the process in industrial systems.

This paper proposes a real-time fault detection and process control based on multi-channel sensor data fusion. The proposed method uses UMLDA-based control chart for monitoring multi-channel data that considers the interrelationship of different signal channels. The proposed method is tested both on simulated and real industrial data and compared with other methods in reference literature. The performance of the proposed method is first evaluated under different scenarios by Monte Carlo simulations. Then the real data are applied to demonstrate the effectiveness of the proposed method. The real data in the case study are from a sensor fusion application in multiple forging operation processes, where multi-channel data are monitored to detect the faults of missing parts and control the process.

The outline of this paper is listed as follows. Section 2 reviews the theoretical background of multilinear algebra and UMLDA method and proposes the real-time fault detection and process control method. In Section 3, the performance of the proposed method is evaluated, and compared with other competitor methods using Monte Carlo simulations. The proposed method is applied to a case study on a multi-operation forging process in Section 4. Finally, this paper is concluded in Section 5.

## II. REAL-TIME FAULT DETECTION AND PROCESS CONTROL METHOD BASED ON FEATURE EXTRACTION OF MULTI-CHANNEL DATA BY UMLDA

In this section, a real-time fault detection and process control method is proposed. The notations relating to the tensor representation and some basic multilinear algebra are presented. The implementations of UMLDA are also reviewed.

### A. Basic definitions and tensor projection

An $N$-way array $\mathcal{X}$ can be denoted as a tensor $\mathcal{X} \in \mathbb{R}^{I_1 \times I_2 \times \ldots \times I_N}$, and each $I_n$ represents the dimension of $n$-mode $n = 1, \ldots, N$. The $n$-mode vector of $\mathcal{X}$ are the vectors whose dimension is $I_n$ got from tensor $\mathcal{X}$ by varying the index $i_n$ while fixing all the other indices. In multilinear algebra, there are two common methods of tensor projection when projecting the tensor into a subspace: the tensor-to-vector projection (TVP) and the tensor-to-tensor projection (TTP).

The TVP consists of multiple Elementary Multilinear Projection (EMPs). The EMPs consist of unit projection vector per mode. Tensor $\mathcal{X} \in \mathbb{R}^{I_1 \times I_2 \times \ldots \times I_N}$ can be projected to a scalar $y$ by an EMP $\{v^{(1)^T}, v^{(2)^T}, \ldots, v^{(N)^T}\}$ as $y = \langle \mathcal{X}, \mathcal{V} \rangle = \langle \mathcal{X}, v^{(1)} \circ v^{(2)} \circ \ldots \circ v^{(N)} \rangle = \mathcal{X} \times_1 v^{(1)^T} \times_2 v^{(2)^T} \ldots \times_N v^{(N)^T}$, $\| v^{(n)} \| = 1$ for $n = 1, 2, \ldots, N$, where $\|\cdot\|$ is the Euclidean norm for vectors, and $\mathcal{V} = v^{(1)} \circ v^{(2)} \circ \ldots \circ v^{(N)}$, and $\circ$ denotes the outer product. The TVP of a tensor $\mathcal{X}$ to a L-dimensional vector $y \in \mathbb{R}^L$ consists of $L$ EMPs $\{v_l^{(1)^T}, v_l^{(2)^T}, \ldots, v_l^{(N)^T}\}, l = 1, \ldots, L$. It can be denoted as

$$\mathcal{X} \times_{n=1}^{N} \{v_l^{(n)^T}, n = 1, \ldots, N\}_{l=1}^{L} \quad (1)$$

in which the $l$th element of $y$ is got by the $l$th EMP as $y(l) = \mathcal{X} \times_1 v_l^{(1)^T} \times_2 v_l^{(2)^T} \ldots \times_N v_l^{(N)^T}$. The TTP is the $n$-mode ($n = 1, 2, \ldots, N$) product of the tensor $\mathcal{X} \in \mathbb{R}^{I_1 \times I_2 \times \ldots \times I_N}$ by the matrix $\mathbf{Z} \in \mathbb{R}^{D_n \times I_n}$ denoted by $\mathcal{X} \times_n \mathbf{Z}$ is a tensor with entries $(\mathcal{X} \times_n \mathbf{Z})(i_1, \ldots, i_{n-1}, d_n, i_{n+1}, \ldots, i_N) = \sum_{i_n} \mathcal{X}(i_1, \ldots, i_N) \cdot \mathbf{Z}(d_n, i_n)$.

### B. UMLDA

The multi-channel data can be denoted as $\mathcal{X} \in \mathbb{R}^{C \times K \times M}$ with M samples (sample index $m = 1, \ldots, M$), $C$ and K are the numbers of channels or sensors (channel or sensor index $c = 1, \ldots, C$) and data points measured in each channel (data index $k = 1, \ldots, K$), respectively. The multi-channel data is also illustrated in Fig.1.

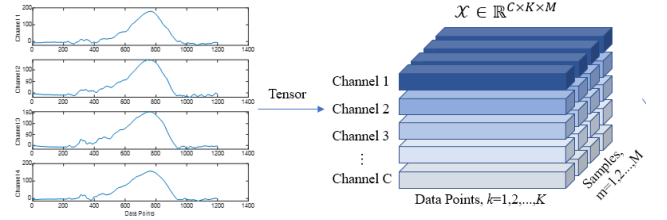

Fig.1. Illustration of multi-channel data

The classical Fisher Discriminant Criterion (FDC) in LDA is defined as the scatter ratio for vector samples, which tries to maximize the between-class distance and minimize the with-class distance simultaneously. The goal of UMLDA is trying to calculate $L$ EMPs $\{v_l^{(n)^T}, n = 1, \ldots, N\}_{l=1}^{L}$ which can maximize the scatter ratio while the features produced are uncorrelated. Let $\{y_{m_l}, m = 1, \ldots, M\}$ denote the $l$th projected (scalar) features, where $y_{m_l} = \mathcal{X}_m \times_1 v_l^{(1)^T} \times_2 v_l^{(2)^T}$ is the projection of the $m$th profile sample using the $l$th EMP. In addition, let $h_l$ denote the set of coordinate vectors. The $m$th projected sample signal using the $l$th EMP is equivalent to the $m$th element of the $l$th coordinate vectors: $h_l(m) = y_{m_l}$. Accordingly, the between-class scatter $S_{B_l}^y$ and the within-class scatter $S_{W_l}^y$ are

$$S_{B_l}^y = \sum_{c=1}^{C} N_c \left(\bar{y}_{c_l} - \bar{y}_l\right)^2 \quad S_{W_l}^y = \sum_{m=1}^{M} \left(y_{m_l} - \bar{y}_{c_{m_l}}\right)^2 \quad (2)$$

Where $C$ is the number of profile classes, $N_c$ is the number of samples for class $c$, $c_m$ is the class label for the $m$th profile sample, $\bar{y}_l = \left(\frac{1}{M}\right) \sum_m y_{m_l} = 0$, and $\bar{y}_{c_l} = \left(\frac{1}{N_c}\right) \sum_{m, c_m = c} y_{m_l}$. The FDC for the $l$th scalar samples $F_l^y = \frac{S_{B_l}^y}{S_{W_l}^y}$. Therefore, the objective function for the $l$th EMP is

$$\{v_l^{(n)^T}, n = 1, 2\} = argmax \, F_l^y \quad (3)$$

$$\text{Subject to} \quad \boldsymbol{v}_l^{(n)T}\boldsymbol{v}_l^{(n)} = 1 \quad \frac{\boldsymbol{h}_l^T \boldsymbol{h}_j}{\|\boldsymbol{h}_l\|\|\boldsymbol{h}_j\|} = \delta_{lj}, \quad l,j = 1,\ldots,L$$

$$\delta_{lj} = \begin{cases} 1, \text{if } l = j \\ 0, \text{otherwise} \end{cases}$$

### C. Proposed real-time fault detection and process control method

After obtaining all the monitoring features, real-time fault detection and process control can be achieved by using control charts. Since there are multiple features, the multivariate Hoteling $T^2$ control chart is used to monitor these features. Let $\boldsymbol{g}_{new} \in \mathbb{R}^J$ denotes the extracted features, then the $T^2$ statistic can be computed as

$$T^2 = (\boldsymbol{g}_{new} - \bar{\boldsymbol{g}})^T \boldsymbol{S}^{-1}(\boldsymbol{g}_{new} - \bar{\boldsymbol{g}}) \quad (4)$$

where the $\bar{\boldsymbol{g}}$ and $\boldsymbol{S}$ are the mean and covariance matrix of feature extracted by in-control training samples. $T^2$ follow an F distribution with J and M-J degrees of freedom. The control limits can be computed by the $(1-\alpha)100^{th}$ percentile of the F distribution, where α is the Type I error. The monitoring procedures are summarized in Fig.2, which consists of Phase I (control chart design) and Phase II (monitoring phase). In Phase I, the model parameters from training data is calculated. Then the control limits are computed with the in-control features until all the training data are in control. In Phase II, the actual monitoring phase, the features and the $T^2$ statistics are computed by the trained model for each multi-channel sample.

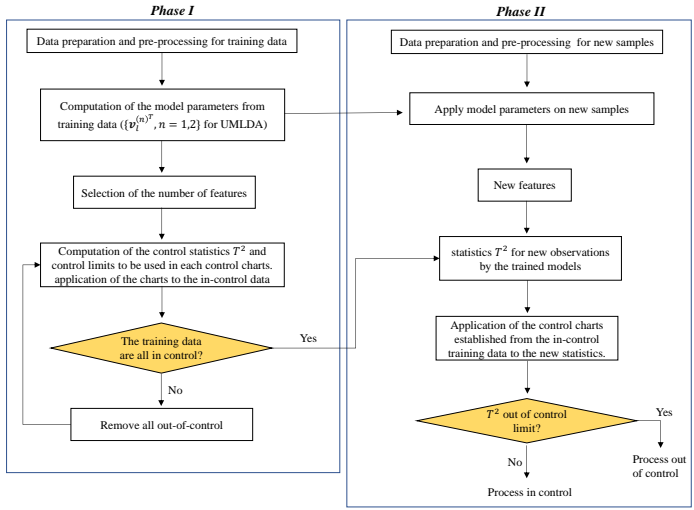

Fig.2. Schematic summary of the required steps to implement the proposed method for process monitoring and fault detection.

## III. PERFORMANCE EVALUATION USING A SIMULATION STUDY

In this section, the Monte Carlo simulation is used to evaluate the performance of the proposed method. The multi-channel profile data is generated with three benchmark signals proposed by Donoho, Johnstone [18] which have been used by different authors [13,19,20].

### A. Multi-channel data generation

The three benchmark signals were chosen for their complex pattern features which are shown in Fig.3. They are called 'Blocks', 'Heavy sine' and 'Bumps', respectively. We denote them as $\boldsymbol{x}_1, \boldsymbol{x}_2$ and $\boldsymbol{x}_3$.

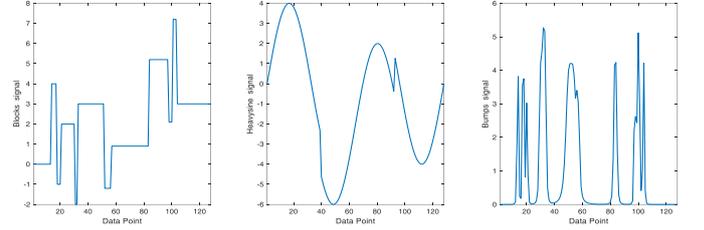

Fig.3. Benchmark signals 'Blocks', 'Heavy sine', and 'Bumps'

In the multi-operation forging process, suppose four sensors are mounted on a forging machine to record 4-channel profiles during each cycle (C = 4), each sensor consists of 128 data points (K = 128). 200 samples (M = 200) are generated in a similar manner as in Grasso et al. [13] which can be denoted as tensor $\mathcal{X} \in \mathbb{R}^{C \times K \times M}$. It can be represented as follows:

$$\begin{aligned}
\mathcal{X}_{1,\cdot,m} &= k_{1.m}\boldsymbol{x}_1 + k_{2.m}\boldsymbol{x}_2 + \varepsilon_{1.m} \\
\mathcal{X}_{2,\cdot,m} &= k_{3.m}\boldsymbol{x}_1^2 + k_{4.m}\boldsymbol{x}_3 + \varepsilon_{2.m} \\
\mathcal{X}_{3,\cdot,m} &= k_{5.m}\boldsymbol{x}_2^2 + k_{6.m}\boldsymbol{x}_3^2 + \varepsilon_{3.m} \\
\mathcal{X}_{4,\cdot,m} &= k_{7.m}\boldsymbol{x}_1\boldsymbol{x}_2 + \varepsilon_{4.m}
\end{aligned} \quad (5)$$

where $\varepsilon_{c.m}$ is the random noise and $\varepsilon_{c.m} \sim N(0,0.5)$, $c = 1,\ldots,4$ $m = 1,2,\ldots,M$. $k_m = [k_{1.m},\ldots,k_{7.m}]^T$ is the model parameter vector and $k_m \sim MVN(\mu_k, \Sigma_k), m = 1,2,\ldots,M$. The model parameter $\mu_k, \Sigma_k$ are set

$$\begin{aligned}
\mu_k &= [0.2,1,1.5,0.5,1,0.7,0.8]^T \\
\Sigma_k &= diag(\sigma_{b_1},\ldots,\sigma_{b_7}) \\
&= diag(0.08,0.015,0.05,0.01,0.09,0.03,0.06)
\end{aligned} \quad (6)$$

Different fault scenarios (out-of-control) are generated from in-control multi-channel data. These out-of-control scenarios are associated with different faults.

(1) Mean shift of the benchmark signals:

$$\begin{aligned}
\boldsymbol{x}_i &= \boldsymbol{x}_i + \delta_1 \boldsymbol{I} \ (i = 1,2,3) \\
\delta_1 &\in \{0.01,0.03,0.05,0.07,0.09\}\sigma_{x_i}
\end{aligned} \quad (7)$$

Where $\sigma_{x_i}$ is the standard deviation of benchmark $\boldsymbol{x}_i$ and $\boldsymbol{I}_{K \times 1}$ is the vector of ones.

(2) Superimposition of a sine function on the benchmark signals:

$$x_i = x_i + \delta_2 \mathbf{y} \quad (i = 1,2,3)$$
$$\delta_2 \in \{0.01, 0.03, 0.05, 0.07, 0.09\}\sigma_{x_i} \quad (8)$$

Where **y** is the sine function with period K and peak-to-peak amplitude equal to 1.

(3) Weak of the part benchmark signals:

$$x_i \to \begin{cases} x_i & 1 \le k \le \frac{5}{8}K \text{ and } \frac{6}{8}K < k \le K \\ x_i - \delta_3 & \frac{5}{8}K < k \le \frac{6}{8}K \end{cases} \quad (i = 1,2,3) \quad (9)$$

$$\delta_3 \in \{0.01, 0.03, 0.05, 0.07, 0.09\}$$

(4) Mean shift of the model parameters:

$$\mu_{k,j} = \mu_{k,j} + \delta_4 \quad (j = 1, \dots, 7)$$
$$\delta_4 \in \{0.1, 0.3, 0.5, 0.7, 0.9\} \quad (10)$$

(5) Standard deviation increase of the random noise:

$$\sigma_{\varepsilon_{c.m}} = \sigma_{\varepsilon_{c.m}} + \delta_5 \quad c = 1, \dots, 4 \quad m = 1, 2, \dots, M$$
$$\delta_5 \in \{0.1, 0.3, 0.5, 0.7, 0.9\} \quad (11)$$

$\sigma_{\varepsilon_{c.m}}$ is the standard deviation of the random noise $\varepsilon_{c.m}$.

A set of 1200 profiles are generated and each class has 200 samples including one in-control condition and five out-of-control conditions (1)-(5). The performance of the proposed method is compared to that of other methods. VPCA is referred to as vectorized-principal component analysis (PCA) [21], which reshapes the tensor data into a vector first and then applies the regular PCA. UMPCA tries to find projections which capture most of the variations in the input high-dimensional data. UMLDA and its developed methods take the class labels into considerations.

*B. Simulation results*

According to the simulation method described above, the multi-channel data are generated and different methods are applied and compared. The performance of all the methods is compared based on the Average Run Length (ARL) with a targeted $\alpha = 0.01$. In order to study the performance with different dimensions of extracted features, 1000 experiments were performed for each projection.

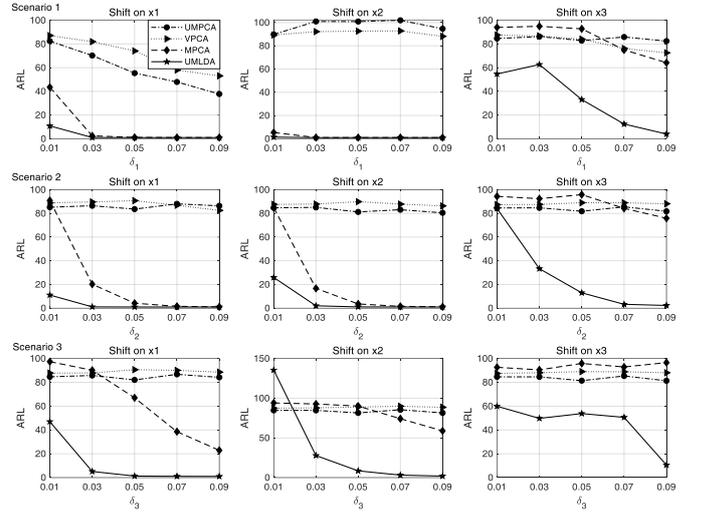

Fig. 4(a) ARLs in simulated Scenarios 1, Scenarios 2 and Scenarios 3

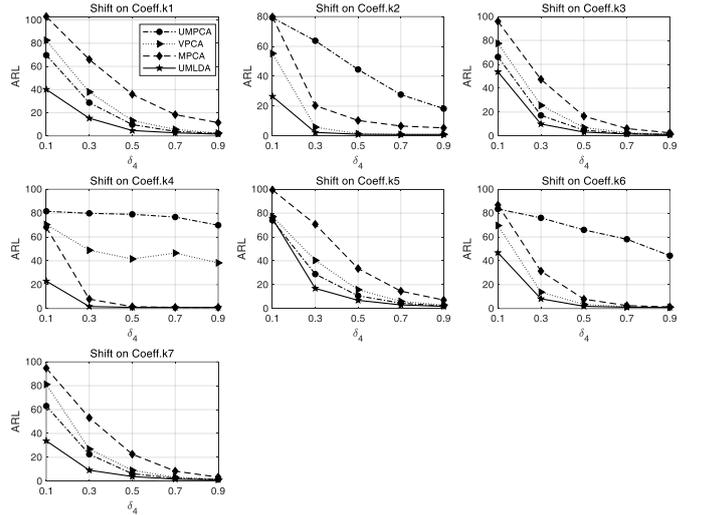

Fig. 4(b) ARLs in simulated Scenarios 4

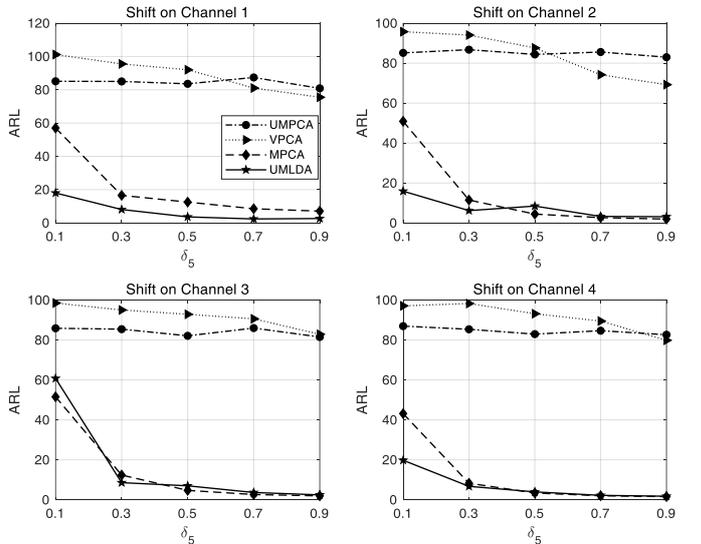

Fig. 4(c) ARLs in simulated Scenarios 5

The plotted results are shown in Fig.4. It is apparent that in most cases the results of UMPCA are worse than methods based on LDA. It is consistent with our understanding that LDA-based methods take class information into classification when extracting features while PCA-based methods only seek projections to maximize the variability. The proposed method exhibits the best performance in the detection in scenario (1), (2) and (3) i.e. when the faults involve the mean shift and superimposition of a sine function on the benchmark signals as well as the weak of the part benchmark signals. In short, when the benchmark signals changes, the UMLDA outweighs all other methods. The MPCA approach becomes a feasible competitor of the UMLDA in the presence of scenarios (4) and (5), i.e. when the deviations involve the distribution of the model parameters and the random noise. However, the MPCA performs a little better than the UMLDA in three on twenty cases (4 channels × 5 δ).

## IV. CASE STUDY

In this section, the proposed method and different other methods are applied to a real-world multi-operation forging process. As illustrated in Fig.5, four sensors are installed on the different uprights of the forging machine, which record the force exerting on dies represented as four-channel profile data. Five different dies are working together to produce a final product which performs five operations in a sequence of (1) performing, (2) blocking, (3) finishing, (4) piercing and (5) trimming. A shape sketch of raw billet, intermediate and final parts after each operation of the selected forging process are shown in Fig.6. The blocking and finishing operations generate large signals because they make significant shape changes on the product while the piercing and trimming operations generate small signals which can also be called weak signals. It's more difficult to detect missing parts in such operations.

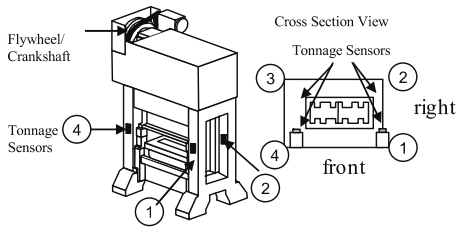

Fig.5. Sensor distributions in the forging machine

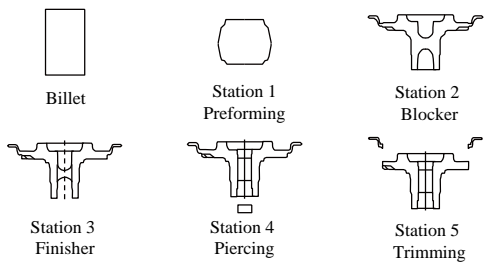

Fig.6. The shape of workpieces at each operation

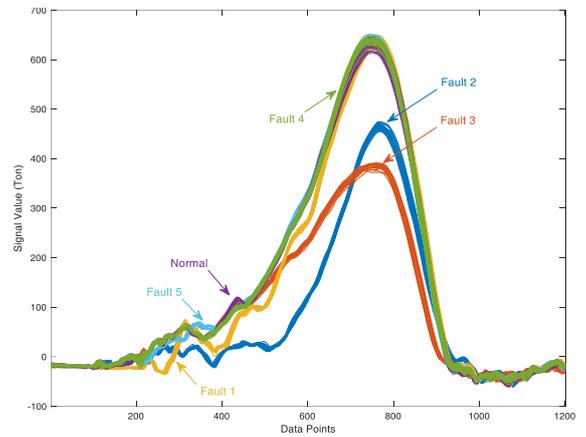

Fig.7. Overlapping samples of aggregated tonnage profiles for normal and missing operations

The proposed method will be used to monitor the four-channel data. Each profile data contains rich information about the product quality which can be used for fault diagnosis. A training multi-channel data set including six groups is collected which contain a normal working production without missing parts in all stations and five faults due to a missing part. Fault $i$ ($i = 1,\ldots,5$) is a faulty condition with a missing part in station $i$ and Fault 0 is the normal working production. The overlapping samples of aggregated multi-channel data for normal working conditions with 308 samples and 69 samples under every five faulty conditions are depicted in Fig.7. The signals partially change at the specific signal segment when a missing part happens. They can be segmented into several parts which specify the working boundary of each operation. This paper will extract features by the corresponding signal segments rather than the whole cycle of the signals in order to increase the detection sensitivity and robustness. Lei et al. [10] added up all channel profiles and applied the vectorized method to the aggregated tonnage profiles to extract features. Paynabar et al. [15] did not break the tensor structure and applied UMPCA to analyze multi-channel profiles that considered the interrelationship of different profile channels. Each profile can be divided into nine segments based on their study which is shown in Table 1. In this paper, segment 4 is chosen to detect all faults from normal working production with the proposed method. The reason why segment 4 is chosen is that the profile under fault 4 condition only changes in segment 4

TABLE 1 Tonnage profile segments (Paynabar et al. [15], Ye et al. [7])

| | Segment | | | | | | | | |
|---|---|---|---|---|---|---|---|---|---|
| | 1 | 2 | 3 | 4 | 5 | 6 | 7 | 8 | 9 |
| Interval | [1,153] | [154,212] | [213,296] | [297,447] | [448,560] | [561,635] | [636,816] | [817,865] | [866,1200] |

As described in simulation, the procedures in Fig.2 are executed step by step for feature extraction and process monitoring. The methods discussed in simulations are applied to these multi-channel data. The in-control and out-of-control samples for each method are plotted on Hoteling $T^2$ charts, which are shown in Fig.8. In addition, the number of all fault samples is 345(69×5). The number of out-of-control samples detected by each method is reported in Table 2. As can be seen in Fig.8 and Table 2, the features extracted by UMLDA lead to better results than features extracted by other methods. This is in accordance with the simulation results. In addition, the average time spent for online monitoring of 1000 samples is also reported in Table 3. This includes the time required for feature extraction and the calculation and comparison of monitoring statistics with control limits. As can be seen from Table 3, even for the slowest method (i.e., UMLDA) it only takes on average 1.4e-3 s to monitor a new sample. This indicates that UMLDA is fast enough for real-time monitoring. It is worth noting that the construction of control charts including the estimation of parameters and projection matrices, and determining control limits is an offline procedure and it is done only once before online monitoring.

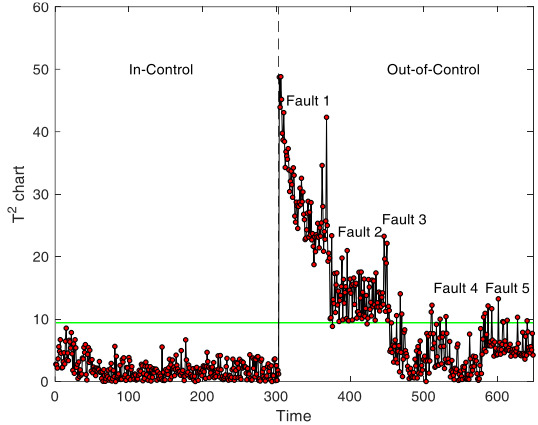
(a)

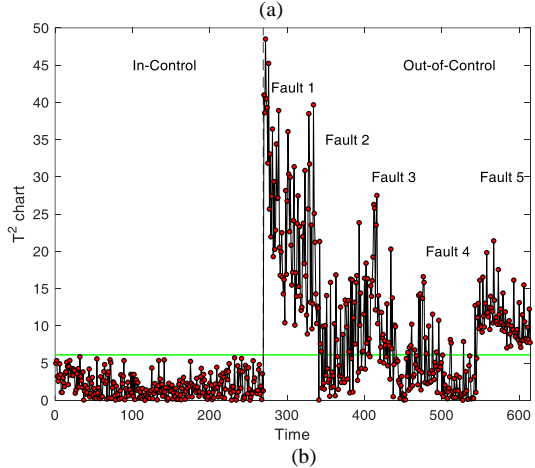
(b)

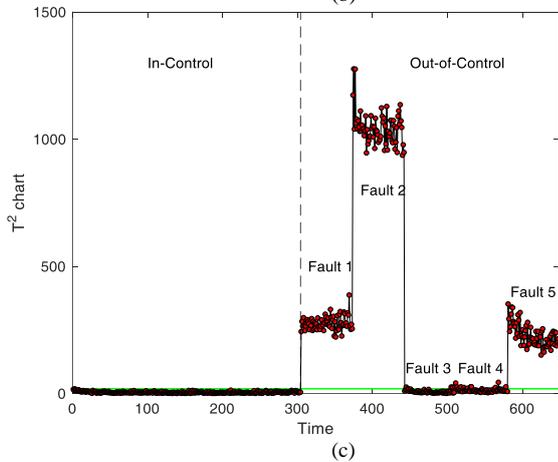
(c)

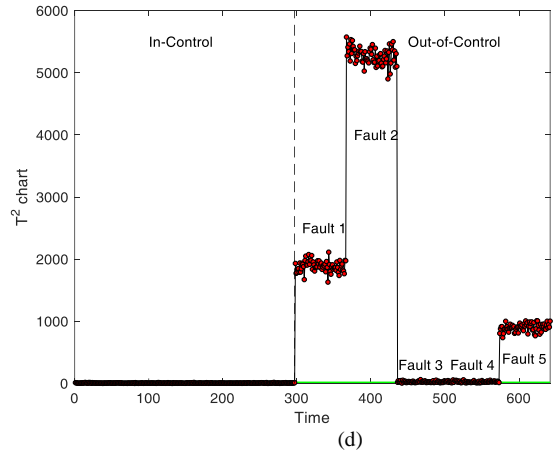
(d)

Fig.8. $T^2$ control chart for process monitoring. (a) VPCA, (b) UMPCA, (c) MPCA and (d) UMLDA

TABLE 2 Number of detected out-of-control samples

|  | VPCA | UMPCA | MPCA | UMLDA |
|---|---|---|---|---|
| Control chart | 163 | 222 | 224 | 311 |

TABLE 3 Average time spent for online monitoring of multi-channel data (s)

|  | VPCA | UMPCA | MPCA | UMLDA |
|---|---|---|---|---|
| Time | 1.3e-3 | 9.5e-5 | 5.4e-4 | 1.4e-3 |

The real-time fault detection and process control of multi-channel data are illustrated in Fig. 9. When the statistics $T^2$ are out of control limits, the process is stopped automatically and immediately for fault diagnosis and protection of the expensive dies.

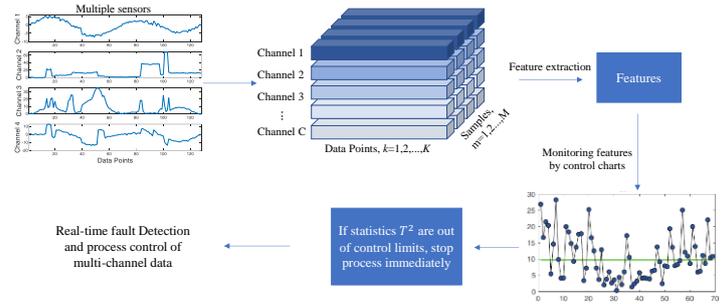

Fig.9. Illustration of real-time monitoring and process control

## V. CONCLUSION

A real-time fault detection and process control method based on multi-channel sensor data fusion is proposed in this paper. The proposed method uses UMLDA to extract features from multi-channel sensor data. The extracted features were combined with multivariate control charts in real-time fault detection and process control. The Monte Carlo simulation was implemented to assess the performance of the proposed method. The simulation result shows that the proposed method has better performance than other competitor methods. These methods were also applied to a real-world case study of a multi-operation forging process for the propose of fault detection and process monitoring. The result shows that the proposed method was more effective than competitor methods for the detection of faults and can be used in real-time fault detection and process control.